\documentclass[letters,usenatbib]{mnras}
\usepackage{graphicx,amssymb,amsmath,times,verbatim}
\usepackage[normalem]{ulem}

\title[$\gamma$-ray QPO in the blazar B2 1520+31]{Detection of a quasi-periodic oscillation in $\gamma$-ray light curve
of the high redshift blazar B2 1520+31}

\author[Gupta et al.]{
Alok C. Gupta$^{1,2}$\thanks{acgupta30@gmail.com},
Ashutosh Tripathi$^{3}$\thanks{ashutosh31tripathi@gmail.com},
Paul J.\ Wiita$^{4}$\thanks{wiitap@tcnj.edu}, 
Pankaj Kushwaha$^{5}$,
\newauthor Zhongli Zhang$^{1}$, Cosimo Bambi$^{3,6}$
\\
\\
$^{1}$Shanghai Astronomical Observatory, Chinese Academy of Sciences, 80 Nandan Road, Shanghai 200030, China \\
$^{2}$Aryabhatta Research Institute of Observational Sciences (ARIES), Manora Peak, Nainital $-$ 263 001, India \\
$^{3}$Center for Field Theory and Particle Physics and Department of Physics, Fudan University, 220 Handan
Road, Shanghai 200433, China \\
$^{4}$Department of Physics, The College of New Jersey, P.O.\ Box 7718, Ewing, NJ 08628-0718, USA \\
$^{5}$Department of Astronomy (IAG-USP), University of Sao Paulo, Sao Paulo 05508-090, Brazil \\
$^{6}$Theoretical Astrophysics, Eberhard-Karls-Universit{\"a}t T{\"u}bingen, D-72076 T{\"u}bingen, Germany
}

\date{Accepted 2018. Received 2018; in original form 2018}

\begin{document}
\label{firstpage}
\pagerange{\pageref{firstpage}--\pageref{lastpage}}
\maketitle

\begin{abstract}
We detected a possible quasi-periodic oscillation (QPO) of $\sim$ 71 days in the 0.1 -- 300 GeV $\gamma$-ray
Fermi-LAT light curve of the high redshift flat spectrum radio quasar B2 1520+31. We identify and confirm that quasi-period by 
Lomb Scargle periodogram (LSP), and weighted wavelet z-transform (WWZ) analyses. Using this QPO period, 
and assuming it originates from accretion-disc fluctuations at the innermost stable circular orbit,  we estimate
the central supermassive black hole mass to range between $\sim 5.4 \times 10^{9}$ M$_{\odot}$ 
for a non-rotating black hole and $\sim 6.0 \times 10^{10}$ M$_{\odot}$ for a maximally rotating black hole. We briefly 
discuss other possible radio-loud active galactic nuclei emission models capable of producing a $\gamma$-ray QPO of 
such a period in a blazar.
\end{abstract}


\section{Introduction}
Active galactic nuclei (AGN) powered by accreting black holes (BHs) with masses of 
10$^{6}$ -- 10$^{10}$ M$_{\odot}$ have several similarities to scaled-up galactic X-ray emitting 
BH binaries. In both BH and neutron star binaries in our and nearby galaxies, the presence of 
quasi-periodic oscillations (QPOs) in the time series data, or light curves, is fairly 
common \citep[e.g.][]{2006ARA&A..44...49R}. But it is quite rare to detect QPOs in the time series data of AGN. 

Blazars are a subclass of radio-loud AGN with their relativistic jets aligned along the observer's 
line of sight. They have been empirically classified further into BL Lac objects (BLLs) and flat 
spectrum radio quasars (FSRQs) based on the strength of optical emission lines, where the former 
show no or very weak ones while the latter have prominent broad lines. All blazars exhibit highly variable fluxes 
across the entire accessible electromagnetic (EM) spectrum from radio to GeV and even TeV $\gamma$-rays 
and on all temporal scales from minutes to decades. This temporal variability is essentially stochastic
\citep[e.g.][]{2017ApJ...849..138K} but there have been occasional claims of QPOs in time series data of blazars 
in different EM bands. Similar to the temporal variability time scales, these QPOs apparently have been seen 
on diverse timescales ranging from a few tens of minutes to hours to days and even years, although many of these
claims are marginal.

Some of the early claims of QPO detections were in the bright blazar OJ 287, where a 15.7 minute periodicity 
in 37 GHz radio observations taken in April 1981 \citep{1985Natur.314..148V} and a 23 minute 
periodicity in optical band observations taken in March 1983 \citep{1985Natur.314..146C}, were argued for.
A quite convincing $\sim$ 11.7 yr QPO was seen using a century long optical data \citep{1996A&A...305L..17S} and
subsequent flares were predicted in terms of a binary BH model \citep{2008Natur.452..851V}. The blazar S5 0716+714 once 
seemed to show a QPO period of $\sim$ 1 day followed by a weaker period of $\sim$ 7 days; these fluctuations were 
present in both optical and radio bands during a coordinated optical and radio monitoring campaign
\citep{1991ApJ...372L..71Q}. On another occasion, optical observations of  S5 0716+714 also indicated a QPO of period of 
$\sim$ 4 days \citep{1996A&A...305...42H}. 
On longer timescales, five optical outbursts during 1995 to 2007 
were suggested to have a quasi-period of $\sim$ 3.0$\pm$0.3 years 
\citep[e.g.][]{2003A&A...402..151R,2006A&A...455..871F,2008AJ....135.1384G}. 
For the blazar PKS 2155$-$304 a possible of QPO of $\sim$ 0.7 day was seen with UV and optical monitoring 
using IUE ({\it International Ultraviolet Explorer}) over five days \citep{1993ApJ...411..614U}. A peculiar blazar,
AO 0235+164, may have shown a QPO of $\sim$ 5.7 years in long term radio band data \citep{2001A&A...377..396R}.       

Over the last decade there have been more claims of detections of QPOs in  several other blazars 
\citep[e.g.][and references therein]{2008ApJ...679..182E,2009ApJ...690..216G,2009A&A...506L..17L,2013MNRAS.436L.114K,2014ApJ...793L...1S,2016ApJ...820...20S,2016AJ....151...54S,2018A&A...615A.118S,2015Natur.518...74G,2015ApJ...813L..41A,2016ApJ...832...47B,2017ApJ...847....7B,2018arXiv180806067B,2017ApJ...847....8L,2017ApJS..229...21X,2017ApJ...835..260Z,2017ApJ...842...10Z,2017ApJ...845...82Z,2018AJ....155...31H}, as well as a few other AGNs of different classes \citep[e.g.][and references therein]{2008Natur.455..369G,2013ApJ...776L..10L,2014ApJS..213...26F,2016ApJ...819L..19P,2017ApJ...849....9Z,2018ApJ...853..193Z,2018A&A...616L...6G}. QPOs  in blazars apparently are occasionally present  on diverse timescales in $\gamma$-ray, 
X-ray, optical, and radio bands, where the  monitoring data has come from a broad range of space and ground based telescopes. Many 
hundreds of light curves with different time resolutions in different EM bands have been analyzed by a variety of 
groups around the globe and QPOs have been only firmly detected in a few light curves of AGN of different sub-classes. We are unaware 
of a claimed detection of  a QPO 
in the same AGN with a nearly similar central period in the same EM band.  
Hence is it a logical conclusion that QPOs in AGNs are both rare and transient in nature.

B2 1520+31 ($\alpha_{2000.0} =$ 15h22m09.99s, $\delta_{2000.0} =+$ 31$^{\circ}$44$^{'}$14.4$^{"}$  
is a high redshift FSRQ located at 
$z = 1.49$ \citep{2012ApJ...748...49S,2017A&A...597A..79P}. This 
blazar was detected in the first 3 months of Fermi-LAT observations and marked as a variable source 
\citep{2009ApJ...700..597A,2010ApJ...722..520A}. It has shown daily activity with $\gamma$-ray flux in the LAT band 
$\geq 10^{-6}$ ph cm$^{-2}$ s$^{-1}$ \citep{2009ATel.2026....1C,2010ATel.3050....1S}. The broadband spectral energy 
distribution (SED) is a typical of FSRQs, with more than an order of magnitude more emission at $\gamma$-ray energies 
than in the optical, so the higher energy bump of the entire spectrum dominates the overall emission \citep{2010ApJ...716...30A}. The 
simultaneous broadband SED of B2 1520+31 has been investigated  \citep{2013MNRAS.436.2170C,2014ApJ...790...45P} and can be 
explained with a one zone emission model. In considering temporal properties \citet{2017ApJ...849..138K} analysed the $\gamma$-ray
Fermi-LAT light curve of this blazar in the energy range 0.1--300 GeV, binned in 3-day intervals. They found that the flux distribution
is log normal,  with a linear relation between flux and intrinsic variability. They suggested that the variability is of a non-linear, 
multiplicative nature and and are consistent with the statistical properties of
 magnetic reconnection powered minijets-in-a-jet model \citep{2012A&A...548A.123B,2012MNRAS.426.1374C}. 

Here we report the first probably QPO detection in the blazar B2 1520+31 with a period of $\sim$ 71 days in 
0.1 -- 300 GeV $\gamma$-ray energies. This is also the first QPO detection in the blazar B2 1520+31 in any EM 
band at any timescale. 

In Section 2 of the manuscript, we briefly describe the $\gamma$-ray Fermi LAT data and our analysis procedure. 
In Section 3 we present  the QPO search methods we employed and the  results of 
those analyses. A discussion and our conclusions are given in Section 4.

\section{Data and reduction}

We downloaded $\rm \gamma-$ray data for B2 1520$+$31 for the period between October 5, 2008 to 
October 5, 2015 (MJD: 54683 -- 57300) from the Large Area Telescope (LAT) on board the space-based Fermi observatory
that was processed through the PASS8 (P8R2) instrument response function. We analyzed the
data with the Fermi Science Tool (v10r0p5) software for photon energies between 100 MeV to
300 GeV. For a given time interval, we first selected the "SOURCE" class registered
events between these energies from a 15$^{\circ}$ circular region of interest (ROI)
centered on the source location (RA: 230.541632, DEC: 31.737328). At the same time,
a maximum zenith angle restriction of 90$^{\circ}$ was applied to avoid the contamination
of $\rm \gamma$-rays from the Earth's limb. The corresponding good time intervals (GTIs)
were generated using the flag "(DATA$\_$QUAL$>$0)\&\&(LAT$\_$CONFIG==1)" which
characterizes the spacecraft operation in Scientific mode. 

Finally, the effect of
selections, cuts, point spread function and presence of point sources were accounted
for in the exposure map, generated on a ROI+10$^{\circ}$ radius. The input model
spectrum XML file of sources within this region was generated using the LAT 3rd
catalog \citep[3FGLgll$\_$psc$\_$v16.fit;][]{2015ApJS..218...23A} which also includes
the contribution of Galactic diffuse and isotropic extra-galactic emission through
the respective emission templates ``gll$\_$iem$\_$v06.fits'' and ``iso$\_$P8R2$\_$SOURCE$\_$V6$\_$v06.txt",
as provided by the LAT Science Team. Finally, the selected events were optimized
against the input model spectrum file and exposure to extract the best fit model
parameters using the python implementation of the ``unbinned likelihood analysis" method
(GTLIKE) provided with the software.

We extracted the light curve of the source
over every 3-day interval by following the above procedures. The optimization over input source model spectrum file was performed
iteratively by removing insignificant sources, measured by Test Statistics (TS) $<$ 0
and freezing the parameters of low TS sources until convergence is reached \citep[e.g.][]{2014ApJ...796...61K}.
All the sources in the model file had the default spectrum from the 3FGL catalog assumed.
Finally, only fluxes $\gtrsim 3\sigma$ defined by a TS of $\gtrsim 9$ were considered,
resulting in a $\sim 97\%$ coverage of the source over the lengthy duration of these observations \citep[e.g.][]{2017ApJ...849..138K}.

\begin{figure}
\centering
\includegraphics[scale=0.43]{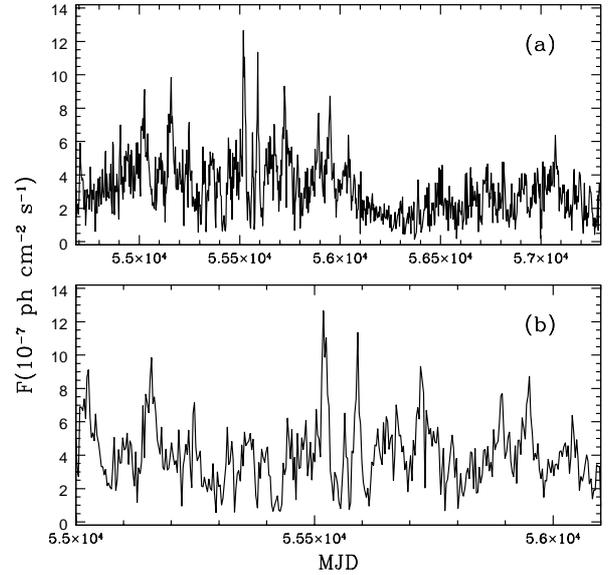}
\vspace*{-0.3in}
\caption{{\bf (a)} 0.1 -- 300 GeV LAT $\gamma$-ray light curve of the blazar B2 1520+31 for
data integration times of three days from August 5, 2008 to October 5, 2015. {\bf (b)} An expanded segment of 
the top panel of the light curve taken between August 5, 2008 and June 22, 2012.} 
\end{figure}

\section{Light curve analysis and results}

The 0.1 -- 300 GeV Fermi LAT $\gamma$-ray light curve of the blazar B2 1520+31, binned in 3 day intervals, for
observations taken from August 5, 2008 to October 5, 2015 is plotted in Fig.\ 1(a).  A visual 
inspection indicated a possible a QPO in the observations made during the first portion of this interval (August 5, 2008 to
June 22, 2012) which are replotted in Fig.\ 1(b). To examine and quantify the possibility of a QPO, we 
analyzed the nearly four-year long light curve data of Fig.\ 1(b) employing the extensively used Lomb-Scargle periodogram (LSP) 
and weighed-wavelet z-transform (WWZ) 
 techniques. In the following subsections we briefly explain these techniques and the QPO 
periods detected by them.

\subsection{Lomb-Scargle periodogram}

The LSP  method is widely used to determine if periodicities are present in 
the data \citep{1976Ap&SS..39..447L,1982ApJ...263..835S} and can be applied to unequally sampled 
data. The method basically involves fitting the sine function throughout the data by using $\chi^2$ 
statistics. It reduces the effect of the noise on the signal and also provides a measure of the significance of any 
periodicity it indicates \citep{2018AJ....155...31H,2017ApJ...835..260Z,2017ApJ...842...10Z,2018ApJ...853..193Z}. 
For more details of our implementation of the LSP, please see \citet[][and references therein]{2018A&A...616L...6G}. 

In Fig. 2, the normalized power of the LSP is plotted against the time period. 
The horizontal line represent the false alarm probability of 0.0001 which corresponds to a nominal 99.99\% 
confidence level. One signal, at the period 
of $70.8_{-2.4}^{+3.7}$ days, reached that significance level. This raises the possibility of there being a true QPO of this period which we  
found to be supported by other methods.

The light curves of AGNs in a range of electromagnetic bands from optical through X-rays and $\gamma$-rays are usually dominated by red noise, which arises from 
stochastic processes in the accretion disks, or in the case of blazars, more likely from jets
\citep[e.g.,][and references therein]{2014ApJS..213...26F,2017ApJS..229...21X,2017ApJ...835..260Z,2017ApJ...845...82Z,2017ApJ...847....7B,2018AJ....155...31H}.
Hence we also employed the REDFIT method 
to fit the data with a  first order auto-regressive (AR1) process 
\citep{2002CG.....28..421S,2014ApJS..213...26F,2017ApJS..229...21X,2018AJ....155...31H,2018A&A...616L...6G}. 
In auto-regressive models the data at a given time are related with previous values through a regressive
relation, and these can involve different numbers of previous values.  In the simplest AR1 model the data point 
at any instance is taken to be related to just the previous one.  This code first computes the time series based on AR1 and generates a theoretical
AR1 spectrum.  It then calculates significance levels based on the $\chi^2$ distribution. 
In Fig.\ 3, the bias corrected power spectrum and a modeled AR1 spectrum are plotted 
against the temporal frequency. We find two peaks which 
are above the displayed 95\% significance level. 
One peak corresponds to the period $70.8_{-0.73}^{+1.83}$ days and the other is 
at $39.33_{-0.56}^{+0.54}$ days. The second peak could be a harmonic of the first peak, which is more 
significant.

\begin{figure}
\centering
\includegraphics[scale=0.5]{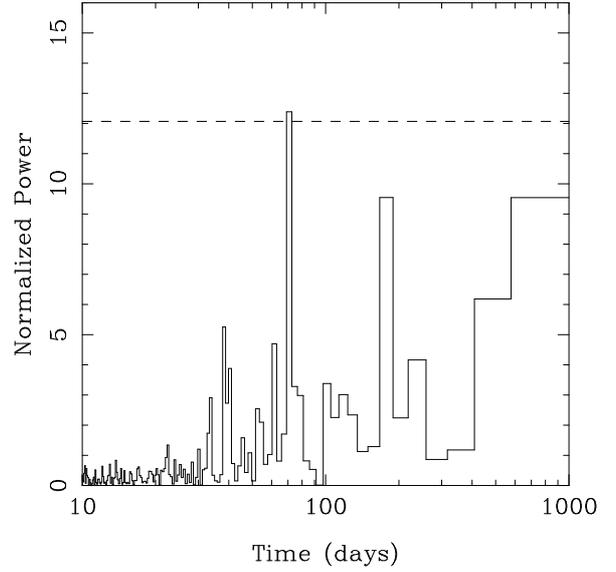} \\
\caption{LSP of the light curve in Fig.\ 1 (b). The dashed line represents a null hypothesis or false alarm probability of $p=0.0001$.}
\end{figure}

\begin{figure}
\centering
\includegraphics[scale=0.5]{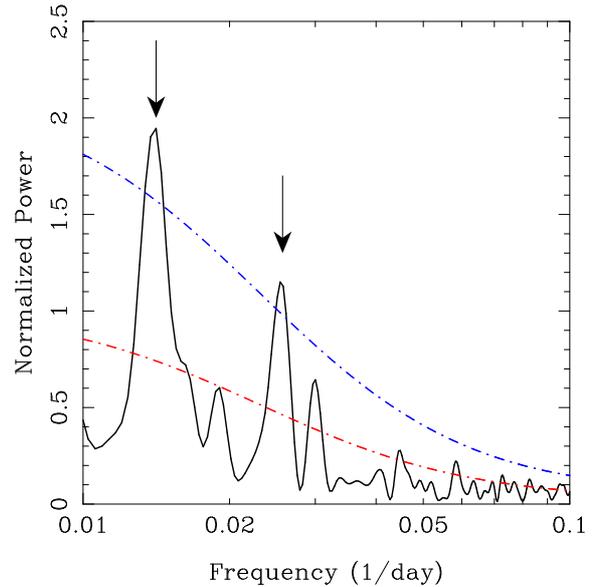}
\caption{Results of the REDFIT method: the black curve represents the bias corrected spectra, the red dot-dashed line 
indicates the computed (AR1) red noise spectrum, and the blue dot-dashed curve shows the 95$\%$ $\chi^{2}$
significance level.}
\end{figure}

\subsection{Wavelet analysis}

Wavelet analyses allow for the determination and estimate of the significance of a period by decomposing the data into time
and frequency domains simultaneously \citep{1998Bulletin}. For more details on this approach, see
\citet[][and references therein]{2018A&A...616L...6G}. We used the WWZ\footnote{https: //www.aavso.org/software-directory} software
to calculate the
weighted wavelet z-transform (WWZ) power for a given time and frequency
\citep[e.g.][and references therein]{2013MNRAS.436L.114K,2016ApJ...832...47B,2017ApJ...847....7B,2018arXiv180806067B,2017ApJ...835..260Z,2017ApJ...842...10Z,2018ApJ...853..193Z}.
To estimate the significance of the signal,
we also calculate the time averaged WWZ power, which gives the strength of the signal at each frequency.

Figure 4 shows the results of our WWZ analysis. The left panel of the figure plots the WWZ determined power.  It illustrates strong  concentrations of power around two periods:  $71.43_{-0.41}^{+0.51}$ days and $178.57_{-3.13}^{+6.25}$. The feature around 71 days is strong 
and persistent throughout most of the observation. The feature a period of around 179 days is of more
moderate strength and is persistent throughout the observation; however, it is very close to 0.5 years and thus has a significant chance of being an observational artifact. The time averaged 
WWZ powers in the right panel of Fig.\ 4 is plotted shows that these periods exceed $3\sigma$ (99.73\%) significance.

\begin{figure*}
\vspace*{-0.7in}
\centering
\includegraphics[angle=90, scale=0.63]{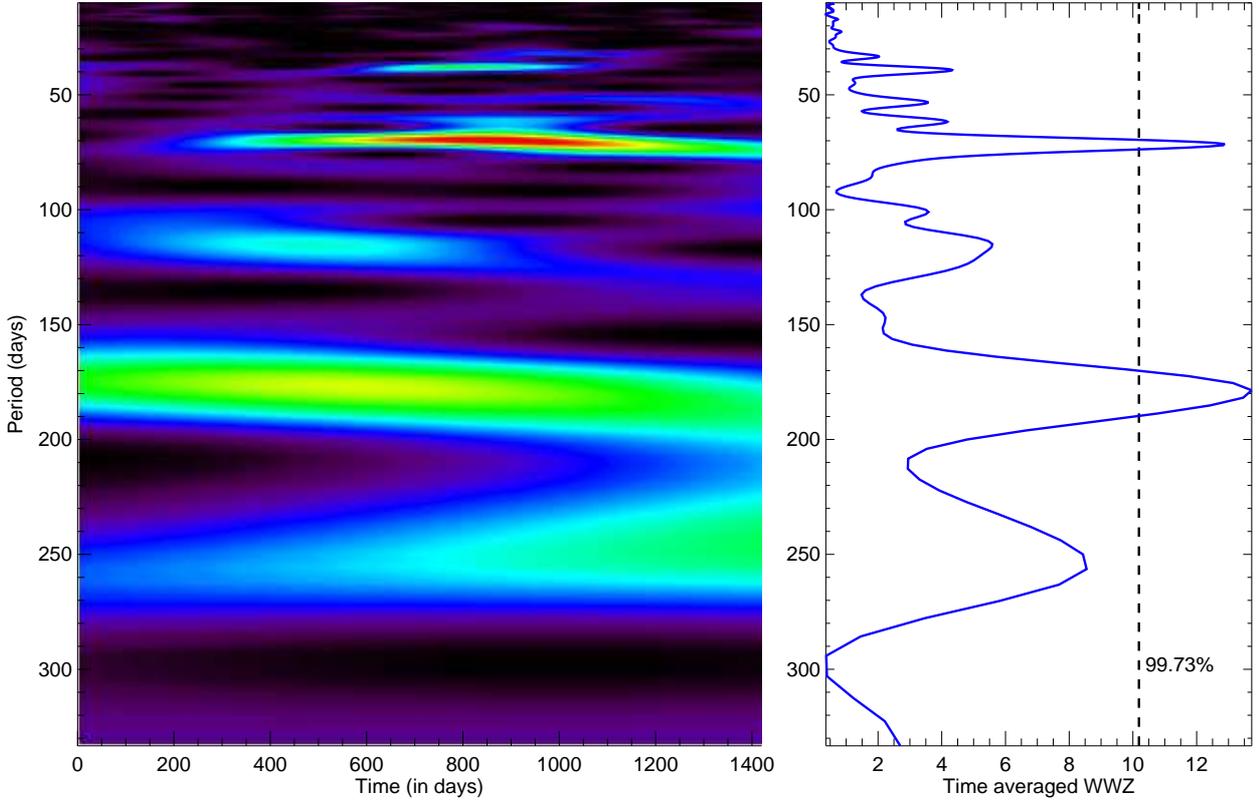}
\caption  {Weighted wavelet z-transform of the light curve presented in Fig.\ 1(b). The left panel shows the
distribution of color-scaled WWZ power (with red most intense and black lowest) in the time-period plane;  the right panel
shows the time-averaged WWZ power (solid blue curve) as a function of period and the $99.73\%$ global significance (dashed black curve).} 
\end{figure*}

\section{Discussion and Conclusion}

We examined the long term 0.1 - 300 GeV energies $\gamma$-ray light curves of four AGNs (the FR I radio galaxy NGC 1275, 
the BL Lac Mrk 421, and the FSRQ, PKS 1510-089) as well as B2 1520+31,  as presented 
 in \citep{2017ApJ...849..138K} to see if they showed any indications of quasi-periodicity. We analyzed these light curves 
 using two techniques, LSP (including REDFIT) and WWZ, which are commonly used for searching for QPOs in AGN time series data 
\citep[][and references therein]{2018A&A...616L...6G}. We found a probable QPO with a period of $\sim$ 71 days in an extended segment 
of the light curve of the FRSQ B2 1520+31, but did not find a QPO in any of the other three AGN.   

In general, blazar emission across the complete EM spectrum is dominated by non-thermal jet emission. This is primarily
because in blazars, the jet is seen at very small angle ($<$ 10$^{\circ}$) the the line of sight to the 
observer \citep{1995PASP..107..803U}. Jet emission will be strongly amplified due to the relativistic 
beaming effect, often overwhelming all the thermal contributions from the AGN and the host galaxies stars. 
But in the FSRQ class of blazars,  relatively efficient accretion disk emission and broad line region (BLR) emission 
lines are present \citep{2011A&A...529A.145D}.
Since B2 1520+31 is a FSRQ, the total emission from this blazar will be expected to have contributions from the accretion disk and the  BLR as well as the jet emission. Many FSRQs show a quasi-thermal excess blue/UV bump above the synchrotron emission in their broad band SEDs 
\citep[e.g.][and references therein]{1999ApJ...521..112P,2004Sci...306..998G,2007A&A...473..819R,2008A&A...491..755R,2009A&A...508..181D,2010ApJ...712..405V,2010ApJ...721.1425A,2011A&A...529A.145D}.  
This portion of the emission from FSRQs is due to both the accretion disk (the so-called ``big blue bump"), 
\citep[e.g.][]{1990MNRAS.246..369L} and the  BLR (the so-called ``little blue bump"), 
\citep[e.g.][]{1985ApJ...288...94W}. 
The contribution of these thermal features will have important consequences in the low-energy part of the SED,
in which this emission could be directly observed.   
Meanwhile, the photons produced by the accretion disk, either directly or through  reprocessing in the BLR or the dusty
torus, are the source of seed photons for the external Compton (EC) mechanism that is often apparently responsible
for  the $\gamma$-ray emission of FSRQs, which comprises the high energy hump of the SED and the $\gamma$-ray fluxes
considered here
\citep[e.g.][and references therein]{2012AJ....143...23G,2017MNRAS.464.2046K,2017MNRAS.472..788G}.

The mass of the central supermassive black hole (SMBH) in an AGN is, along with the accretion rate and
efficiency of mass to energy conversion, one of the most important quantities to characterize.
The most accurate, or primary, black hole mass estimation methods include stellar and gas kinematics and reverberation mapping
\citep[e.g.][]{2004ASPC..311...69V}. All these methods require high spatial resolution spectroscopy data from the 
host galaxy and/or higher-ionization emission lines and are not applicable to most blazars. 
The BLL class of blazars  have essentially featureless spectra, so, primary methods can not be used. But in 
the case of FSRQs, prominent emission lines are present, so we can use the method \citep{2006ApJ...641..689V}. 

An alternative way to estimate  the SMBH mass of an AGN comes from using the period of a detected QPO if we assume the QPO is related to the
orbital timescale of a hot spot, spiral shocks, or other non-axisymmetric phenomena in the innermost portion 
of the rotating accretion disk \citep[e.g.][]{1991A&A...246...21Z,1993ApJ...406..420M,1993ApJ...411..602C,2012MNRAS.423.3083M}. 
Using this assumption for the origin of a QPO, one has an expression for the SMBH mass, $M$ \citep{2009ApJ...690..216G} 
   \begin{equation}
\frac {M} {M_{\odot}} = {\frac {3.23 \times 10^4 ~ P} {(r^{3/2} + a)(1+z)}},
\end{equation}
in terms of the QPO period $P$ in seconds and the radius of this source zone, $r$ (in units of $GM/c^2$), and SMBH spin parameter $a$.
The range of nominal masses of the SMBH with such a QPO source can be evaluated in this fashion for perturbations at the innermost stable circular orbit for a
Schwarzschild BH (with $r = 6.0$ and $a = 0$), 
and for a maximal Kerr BH (with $r = 1.2$ and $a = 0.9982$) \citep{2009ApJ...690..216G}.

In the case of FSRQ B2 1520+31, using equation (1) for the period of 71 days, we get an SMBH mass estimate
of $5.41 \times 10^{9}$ M$_{\odot}$ for the Schwarzschild limit  and $6.02 \times 10^{10}$ M$_{\odot}$
for the maximal Kerr limit.  Even the former estimate is very large while the  latter exceeds essentially all other SMBH mass estimates
\citep[e.g.][]{2004ApJ...611..761D,2012MNRAS.427...77V,2015Natur.518..512W,2015ApJ...799..189Z,2015MNRAS.450L..34G}, so attributing this apparently detected QPO to emission directly reflecting a transient non-axisymmetric accretion structure is rather unlikely, particularly for the high energy emission that would not directly emerge from the disk.   If the  portion of the disk taken to be responsible for a QPO is further out than the innermost portions assumed above,
then the estimated mass decreases, perhaps to more reasonable values.

Nonetheless, it is also a prior more likely that  the detected QPO in any blazar is related to the jet emission and not directly to that of the accretion disk. If the jet precesses or has an internal helical structure, which is certainly plausible in blazars \citep[e.g.][]{1992A&A...255...59C,1999A&A...347...30V,2004ApJ...615L...5R,2015ApJ...805...91M}, then as shocks advance along the helical structure of the jet or as the jet precesses or twists, quasi-periodic flux variations  would arise from variations in the Doppler boosting factor as seen by the observer.  \citet{2015ApJ...813L..41A}, in describing a possible roughly 2 year QPO in the $\gamma$-ray (and other band as well) emission from the blazar PG 1553+113, nicely summarize several possibilities along these lines.  For instance, Lense-Thirring precession of the disk \citep[e.g.][]{1972PhRvD...5..814W} could modulate the direction of the jet \citep[e.g.][]{2009ApJ...693..771F}.  

Another clear way to induce jet precession is for the AGN to be part of a binary SMBH system \citep[e.g.,][]{1980Natur.287..307B,2008Natur.452..851V,2015Natur.518...74G}, but these orbits are most likely to produce physical periods in the jets exceeding 1 year \citep[e.g.][]{2007Ap&SS.309..271R} and several candidate $\gamma$-ray QPOs with periods longer than that have recently have been discussed in this framework  \citep[e.g.][]{2015ApJ...813L..41A, 2016AJ....151...54S,2016ApJ...820...20S,2017ApJ...835..260Z,2017ApJ...842...10Z,2017ApJ...845...82Z}  However, as noted by \citep{2004ApJ...615L...5R}, the observed periods could be substantially shorter for sufficiently well aligned jets and large enough Lorentz factors.  This scenario was shown to be quite reasonable for an apparent QPO in the BLL PKS 2247$-$131 of an even shorter observed oscillation at around $\sim 34$ days (Zhou et al. 2018). 

Unfortunately, B2 1520+31 was not the subject of frequent VLBI measurements, nor was there good optical monitoring of this blazar between 2008 and 2012. 
Without a good measurement of the jet Lorentz factor and the angle to our line of sight of the center of the jet of  B2 1520+31 it is not possible to reasonably constrain the parameters of any of these jet based models to produce the variations in Doppler factors that would be required to yield the high amplitude apparent $\sim 71$d QPO we have found.

\section*{Acknowledgements}

This research has used data, software, and web tools of High Energy Astrophysics Science 
Archive Research Center (HEASARC), maintained by NASA's Goddard Space Flight Center.

ACG is grateful for hospitality at SHAO during his visit. AT acknowledges support from 
the China Scholarship Council (CSC), Grant No. 2016GXZR89. PK acknowledges support from 
FAPESP grant no. 2015/13933-0. ZZ is thankful for support from the CAS Hundred-Talented 
program (Y787081009). CB was supported by the National Natural Science Foundation of China 
(Grant No. U1531117), Fudan University (Grant No. IDH1512060), and the Alexander von Humboldt
Foundation.

% Don't change these lines
\bsp	% typesetting comment
\label{lastpage}
\end{document}